\documentclass[12pt]{article}

\usepackage{graphics}
\usepackage{graphicx}
\usepackage{amssymb}
\usepackage{amsmath}
\usepackage{amsfonts}
\usepackage{cite}

\begin{document}

\begin{titlepage}
\begin{center}

{\Large \bf{Theoretical study of hydrogenated tetrahedral aluminum clusters}
}

\vskip .45in

{\large
Kazuhide Ichikawa$^1$, Yuji Ikeda$^1$, Ayumu Wagatsuma$^1$, \\ 
Kouhei Watanabe$^1$, Pawe{\l}  Szarek$^2$ and Akitomo Tachibana$^{*1}$}

\vskip .45in

{\em
$^1$Department of Micro Engineering, Kyoto University, Kyoto 606-8501, Japan\\
$^2$Wroc{\l}aw University of Technology, Institute of Physical and Theoretical Chemistry, Wybrze\.ze Wyspia\'nskiego 27, 50-370 Wroc{\l}aw, Poland
}

\vskip .45in
{\tt E-mail: akitomo@scl.kyoto-u.ac.jp}

\end{center}

\vskip .4in

\date{\today}

\begin{abstract}
We report on the structures of aluminum hydrides derived from a tetrahedral aluminum (Al$_4$) cluster using {\it ab initio} quantum chemical calculation. 
Our calculation of binding energies of the aluminum hydrides reveals that stability of these hydrides increases as more hydrogen atoms are adsorbed, while stability of Al--H bonds decreases.
We also analyze and discuss the chemical bonds of those clusters by using recently developed method based on the electronic stress tensor.
\end{abstract}

{wave function analysis; theory of chemical bond; stress tensor; hydrogenated aluminum cluster}

\end{titlepage}

\setcounter{page}{1}


\section{Introduction} \label{sec:intro}

Recently, hydrogen is paid much attention as new primary energy source because of the 
depletion of fossil fuels and environmental issues such as global warming. 
There are three basic research challenges -- production, storage, application -- for a 
``hydrogen economy". Our study in this paper is related to hydrogen storage. As is 
emphasized in the recent reports on basic research challenges for hydrogen storage, its high 
efficiency is a key factor in enabling the success of the hydrogen economy \cite{Read2006,ACIL2003}. Hydrogen 
storage systems must exhibit following properties: appropriate thermodynamics, fast kinetics, 
high storage capacity (more than 10 wt \%), effective heat transfers, high volumetric densities, 
long cycle lifetime, safety under normal use \cite{ANL2003}. To implement these properties, solid-state 
storage is useful. Metal hydrides, chemical storage materials, nanostructured materials are 
well known for effective solid-state storage systems. 

Among them, we investigate aluminum hydrides in the form of aluminum clusters. More specifically, we deal with an aluminum 
tetrahedral cage (Al$_4$) and its hydrides which were recently observed in experiment and 
confirmed to have enhanced stability \cite{Li2007,Grubisic2007}. 
Although the clusters found in Refs.~\cite{Li2007,Grubisic2007} are produced by vaporizing aluminum metal in
hydrogen gas and do not lead to immediate application for hydrogen storage material at this stage, 
it would be useful to study their properties theoretically to further explore possible connection with 
more realistic materials for hydrogen storage. 
It is also interesting to study the tetrahedral cage in the 
aspect that it is the fundamental structure of an aluminum icosahedral cage (Al$_{12}$) \cite{Goldberg2007}. 
In this paper, we report on the structures of aluminum hydrides which can be constructed by adding H$_2$ molecules to an Al$_4$ tetrahedral cluster.

We also evaluate and discuss the chemical properties of aluminum clusters and hydrogen adsorption by 
using a novel method of the electronic stress tensor based on the Regional Density Functional Theory (RDFT) and Rigged Quantum Electrodynamics (RQED)
\cite{Tachibana1999,Tachibana2001,Tachibana2002a,Tachibana2002b,Tachibana2003,Tachibana2004,Tachibana2005,Szarek2007,Szarek2008,Szarek2009a,Ichikawa2009a,Ichikawa2009b,Tachibana2010}.

This paper is organized as follows. In the next section, we briefly explain our quantum chemical computation method. We also describe our analysis method based on the RDFT  and the RQED, and in particular we define our bond orders and regional chemical potential. In Sec.~\ref{sec:results}, we discuss our results. Sec.~\ref{sec:structure} shows structures of hydrogenated Al$_4$ tetrahedral clusters and discuss their binding energies. In Sec.~\ref{sec:bo}, we analyze the structures using the electronic stress tensor and our bond orders. 
We summarize our paper in Sec.~\ref{sec:summary}.

\section{Theory and calculation methods} \label{sec:calc}
\subsection{Ab initio electronic structure calculation}
We perform ab initio quantum chemical calculation for several clusters of aluminum 
hydrides derived from an Al$_4$ tetrahedral cage.
In this work, calculations are performed by {\sc{Gaussian03}} program package \cite{Gaussian03} using density functional theory (DFT) 
with Perdew-Wang 1991 exchange and correlation function (PW91) \cite{Hamann1989,Perdew1992}. 
The split-valence triple-zeta 6-311G** basis set \cite{Binning1990,McGrath1991,Curtiss1995} with polarization functions has been used.
Optimization was performed without imposing symmetry.
The visualization is done using Visual Molecular Dynamics (VMD) \cite{VMD} and 
PyMOL Molecular Viewer programs \cite{PyMOL}. 

\subsection{RDFT analysis} \label{sec:RDFT}
In the following section, we use quantities derived from the electronic stress tensor to analyze electronic structures of hydrogenated Al$_4$ clusters. This method based on RDFT and RQED\cite{Tachibana1999,Tachibana2001,Tachibana2002a,Tachibana2002b,Tachibana2003,Tachibana2004,Tachibana2005,Tachibana2010} include useful quantities to investigate chemical bonding and reactivity such as new definition of bond order and regional chemical potential \cite{Szarek2007,Szarek2008,Szarek2009a}. We briefly describe them below.
(For other studies of quantum systems with the stress tensor in a slightly different context, see Refs.~\cite{Bader1980,Nielsen1983,Nielsen1985,Folland1986a,Folland1986b,Godfrey1988,Filippetti2000,Pendas2002,Rogers2002,Morante2006,Tao2008,Ayers2009}.)

The basic quantity in this analysis is the electronic stress tensor density $\overleftrightarrow{\tau}^{S}(\vec{r})$ whose components are given by
\begin{eqnarray} 
\tau^{Skl}(\vec{r}) &=& \frac{\hbar^2}{4m}\sum_i \nu_i
\Bigg[\psi^*_i(\vec{r})\frac{\partial^2\psi_i(\vec{r})}{\partial x^k \partial x^l}-\frac{\partial\psi^*_i(\vec{r})}{\partial x^k} \frac{\partial\psi_i(\vec{r})}{\partial x^l} \nonumber\\
& &+\frac{\partial^2 \psi^*_i(\vec{r})}{\partial x^k \partial x^l}\psi_i(\vec{r}) -\frac{\partial \psi^*_i(\vec{r})}{\partial x^l}\frac{\partial \psi_i(\vec{r})}{\partial x^k}\Bigg],
\label{eq:stress}
\end{eqnarray}
where $\{k, l\} = \{1, 2, 3\}$, $m$ is the electron mass, $\psi_i(\vec{r})$ is the $i$th natural orbital and $\nu_i$ is its occupation number.

Taking a trace of $\overleftrightarrow{\tau}^{S}(\vec{r})$ can define energy density of the quantum system at each point in space. The energy density $\varepsilon_\tau^S(\vec{r})$ is given by 
\begin{eqnarray}
\varepsilon_\tau^S(\vec{r}) = \frac{1}{2} \sum_{k=1}^3 \tau^{Skk}(\vec{r}).
\end{eqnarray}
We note that, by using the virial theorem, integration of $\varepsilon_\tau^S(\vec{r})$ over whole space gives usual total energy $E$ of the system: $\int \varepsilon_\tau^S(\vec{r}) d\vec{r} = E$.

Regional chemical potential $\mu_R$ \cite{Tachibana1999} is calculated approximately using $\varepsilon_\tau^S(\vec{r})$ \cite{Szarek2007}. 
\begin{eqnarray}
\mu_R = \frac{\partial E_R}{\partial N_R} \approx \frac{\varepsilon_\tau^S(\vec{r})}{n(\vec{r})}, \label{eq:mu}
\end{eqnarray}
where $n(\vec{r})$ is the ordinary electron density at $\vec{r}$. Since electrons tend to move from high $\mu_R$ region to low $\mu_R$ region, the distribution of $\mu_R$ maps the chemical reactivity. 

Now, we define bond orders as $\varepsilon_\tau^S(\vec{r})$ or $\mu_R$ at ``Lagrange point" \cite{Szarek2007}. The Lagrange point $\vec{r}_L$ is the point where the tension density $\vec{\tau}^S(\vec{r})$ given by the divergence of the stress tensor 
\begin{eqnarray} 
\tau^{S k}(\vec{r})&=&  \sum_l \partial_l  \tau^{Skl}(\vec{r}) \nonumber \\
&=&\frac{\hbar^2}{4m}\sum_i \nu_i
\Bigg[\psi^*_i(\vec{r})\frac{\partial \Delta\psi_i(\vec{r})}{\partial x^k}-\frac{\partial\psi^*_i(\vec{r})}{\partial x^k} \Delta\psi_i(\vec{r}) \nonumber\\
& &+\frac{\partial \Delta\psi^*_i(\vec{r})}{\partial x^k}\psi_i(\vec{r}) -\Delta \psi^*_i(\vec{r}) \frac{\partial \psi_i(\vec{r})}{\partial x^k}\Bigg],
\end{eqnarray}
vanishes. Namely, $\tau^{S k}(\vec{r}_L)=0$. $\vec{\tau}^S(\vec{r})$ is the expectation value of the tension density operator $\Hat{\vec{\tau}}^S(\vec{r})$, which cancels the Lorentz force density operator $\Hat{\vec{L}}(\vec{r})$ in the equation of motion for stationary state \cite{Tachibana2003}. Therefore, we see that $\vec{\tau}^S(\vec{r})$ expresses purely quantum mechanical effect and it has been proposed that this stationary point might characterize chemical bonding \cite{Szarek2007}. Then, our newly defined bond orders are
\begin{eqnarray}
b_\varepsilon = \frac{\varepsilon^S_{\tau {\rm AB}}(\vec{r}_L)}{\varepsilon^S_{\tau {\rm HH}}(\vec{r}_L)}, \label{eq:be}
\end{eqnarray}
and
\begin{eqnarray}
b_\mu = \frac{\varepsilon^S_{\tau {\rm AB}}(\vec{r}_L) / n_{\rm AB}(\vec{r}_L)}{\varepsilon^S_{\tau {\rm HH}}(\vec{r}_L) / n_{\rm HH}(\vec{r}_L)}. \label{eq:bmu}
\end{eqnarray}
One should note normalization by the respective values of  a H$_2$ molecule calculated at the same level of theory (including method and basis set).

We use Molecular Regional DFT (MRDFT) package \cite{MRDFTv3} to compute these quantities introduced in this section.

\section{Results and discussion} \label{sec:results}

\subsection{Structures and stability} \label{sec:structure}

\begin{figure*}
\begin{center}
\includegraphics[scale=0.7]{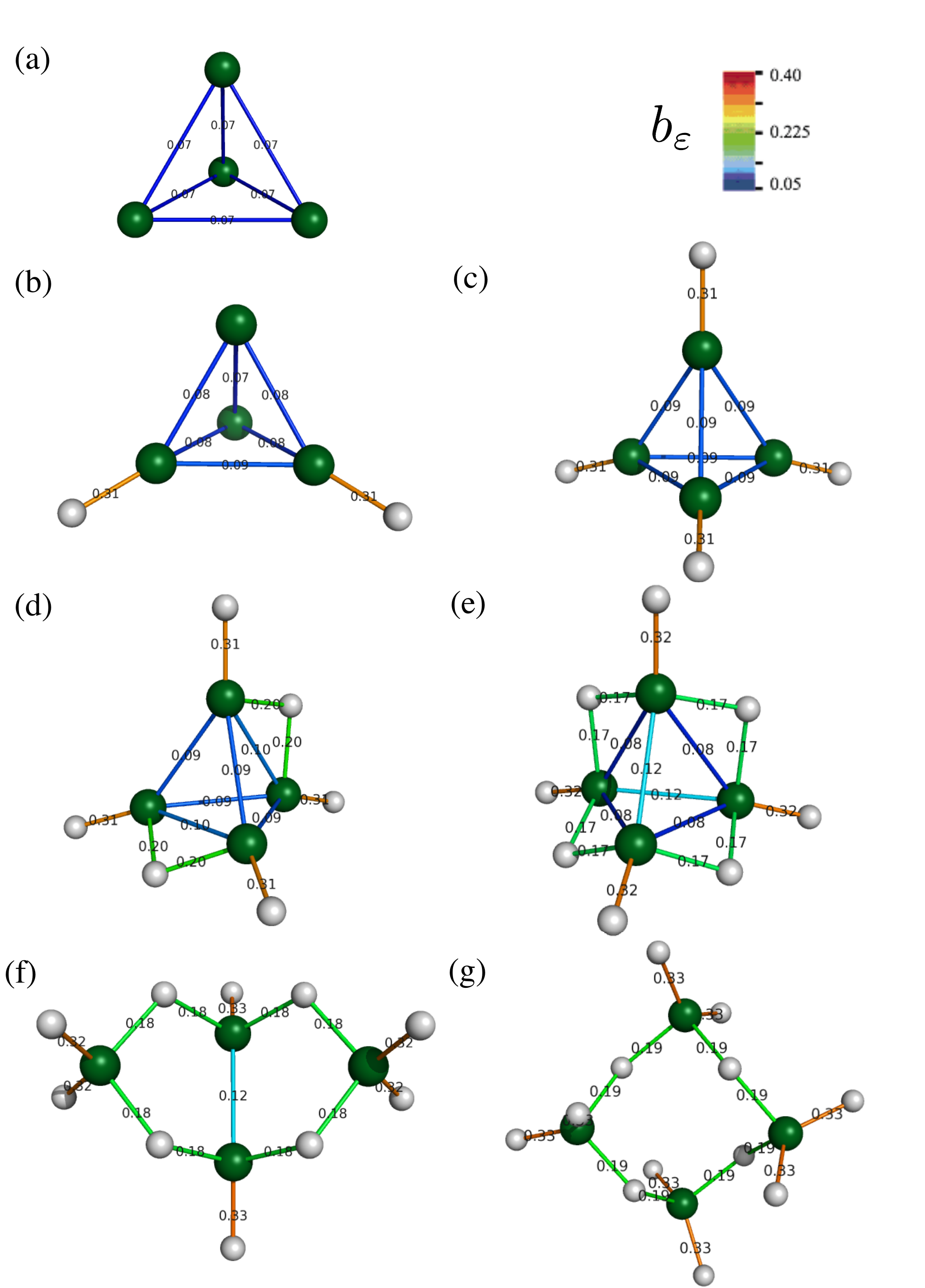}
\caption{Optimized structures for Al$_4$ tetrahedral cage and its hydrides Al$_4$H$_n$ ($n=2, 4, 6, 8, 10, 12$). The bonds are drawn at which Lagrange points are found and our energy density based bond order $b_\varepsilon$ is shown.  }
\label{fig:bo}
\end{center}
\end{figure*}

The bare tetrahedral optimized structure of Al$_4$ is shown in Fig.~\ref{fig:bo} (a). We found that 
all the six Al---Al bonds have an equal length of 2.74\,{\AA} to a great accuracy. Al$_4$ is considered to 
have a structure very close to a regular tetrahedron. 
We note that this regular tetrahedral structure is stable only for high spin state (multiplicity $=5$). 
Also it should be noted that the global minimum of Al$_4$ cluster is planar rhombus with multiplicity 3 \cite{Jones1991,Kawamura2001},
which has lower energy by 0.5 eV.
Below, we investigate the structures when hydrogens adsorbs to this tetrahedral structure.

We first considered the adsorption of two hydrogen atoms. We examined many combinations of adsorption sites 
and multiplicities 1, 3 and 5.
We found that Al$_4$H$_2$ with two hydrogen atoms at terminal sites as shown in Fig.~\ref{fig:bo} (b) and with multiplicity 3, has the lowest energy. 

We further added hydrogen atoms to this structure. Fig.~\ref{fig:bo} (c)-(g) show structures of most stable 
isomer of Al$_4$H$_n$ ($n=4, 6, 8, 10$ and 12) we have obtained.
 We have tried to 
adsorb more hydrogen, without success. We could not find a stable structure for Al$_4$H$_{14}$. Thus, 
we believe Al$_4$H$_{12}$ is the saturated structure of the Al$_4$ cluster. 

Here, we comment on the comparison with the structures which were reported in literatures. 
The structure of Al$_4$H$_4$, Fig.~\ref{fig:bo} (c), is consistent with Ref.~\cite{Moc2006,Grubisic2007} and
Al$_4$H$_6$, Fig.~\ref{fig:bo} (d), is consistent with Ref.~\cite{Grubisic2007}.
As for Al$_4$H$_{12}$, there are several literatures which investigated stable structures for this as a tetramer of alane Al\,H$_3$ \cite{Shen1993,Willis2000,Kawamura2003,Moc2006}. This structure also attracts interest because of its high hydrogen storage capacity of 10.0 wt\%, which exceeds the target value of a hydrogen storage system 
specified by some technical report \cite{ANL2003}. 
Ref.~\cite{Moc2006} has pointed out that structures in Refs.~\cite{Shen1993} and \cite{Willis2000} are not stable in light of recent quantum chemical computation
and reported a structure with S$_4$ symmetry as the global minimum. Our results, Fig.~\ref{fig:bo} (g), agrees with the structure found in Ref.~\cite{Moc2006}.
Ref.~\cite{Moc2006} also reported stable structures of Al$_4$H$_{10}$, including the structure we have found as Fig.~\ref{fig:bo} (f).
They have shown that there is a structure with lower energy but since that structure is chain-like, we adopt the structure of Fig.~\ref{fig:bo} (f)
as the one which is derived by adding hydrogen atoms to Fig.~\ref{fig:bo} (e).

In order to investigate the stability of these structures, we here define two types of binding energies (B.E.).
The total B.E. of Al$_4$H$_m$ is 
\begin{eqnarray}
\Delta E_{\rm total} = E({\rm Al}_4 {\rm H}_m) - [4 E({\rm Al}) +m E({\rm H}) ],
\end{eqnarray}
where $E({\rm X})$ is the energy of X. $\Delta E_{\rm total}$ represents the sum of the strength of all the bonds existing in the molecule. 
The average B.E. of H atoms is defined as
\begin{eqnarray}
\Delta E_{\rm H} = \frac{1}{m} \left\{  E({\rm Al}_4 {\rm H}_m) - [E({\rm Al}_4) +m E({\rm H}) ]  \right\},
\end{eqnarray}
which represents the strength of Al---H bond per one hydrogen atom.
We use the structure of Fig.~\ref{fig:bo} (a) to calculate $E({\rm Al}_4)$.

\begin{figure}
\begin{center}
\includegraphics[scale=0.55]{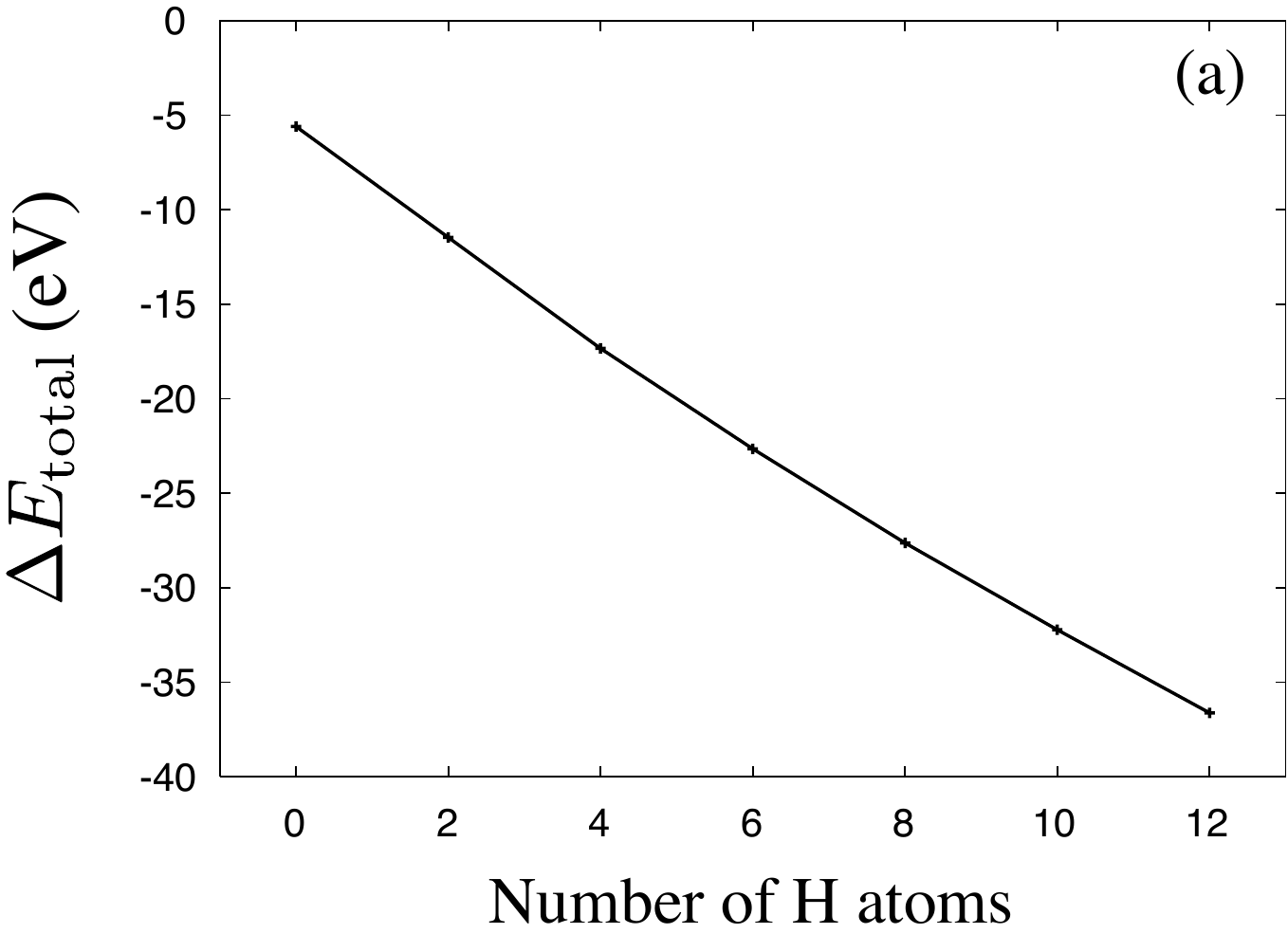}
\includegraphics[scale=0.55]{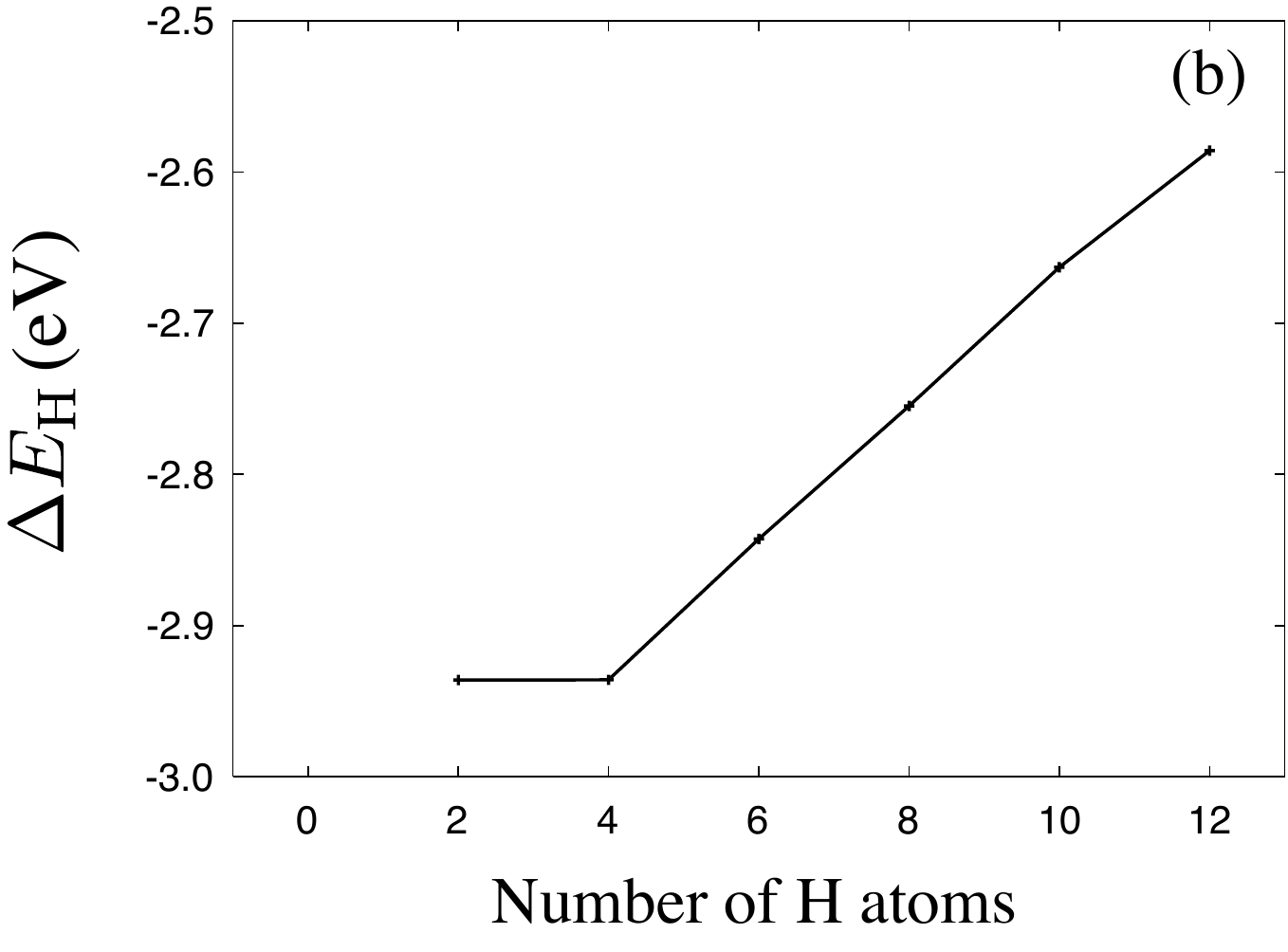}
\caption{Total B.E. $\Delta E_{\rm total}$ (panel(a)) and average B.E. of H atoms $\Delta E_{\rm H}$ (panel (b)) of Al$_4$H$_n$ ($n=0, 2, 4, 6, 8, 10$ and 12). }
\label{fig:BE_Al4Hn}
\end{center}
\end{figure}

The results for these two types of binding energy are summarized in Table \ref{tab:Al4Hn} and Fig.~\ref{fig:BE_Al4Hn}.
As is shown in Fig.~\ref{fig:BE_Al4Hn} (a), $\Delta E_{\rm total}$ decreases as more hydrogens are adsorbed. 
This indicates that adsorption of hydrogen stabilizes the cluster. 
On the other hand, as we plot in Fig.~\ref{fig:BE_Al4Hn} (b), $\Delta E_{\rm H}$ is almost same for Al$_4$H$_2$ and Al$_4$H$_4$ but
increases as more hydrogens are adsorbed. 
This is considered to be due to the increased hydrogens at bridge sites for larger clusters than Al$_4$H$_6$. 
It should be stressed that lower total B.E. and higher average B.E. of H atoms are favorable for hydrogen storage systems. 
This means that hydrogen atoms turn into  
state such that they are easily desorbed, while clusters become more stable as hydrogen are adsorbed.

\subsection{Stress tensor analysis of chemical bond} \label{sec:bo}

In the previous subsection, we have shown structures of aluminum hydrides derived from the Al$_4$ 
tetrahedral cage and discussed the stability of these clusters by the usual binding energy. In 
this section, we discuss their chemical bonds using RDFT analysis introduced in Sec.~\ref{sec:RDFT}.

\begin{figure}
\begin{center}
\includegraphics[scale=0.6]{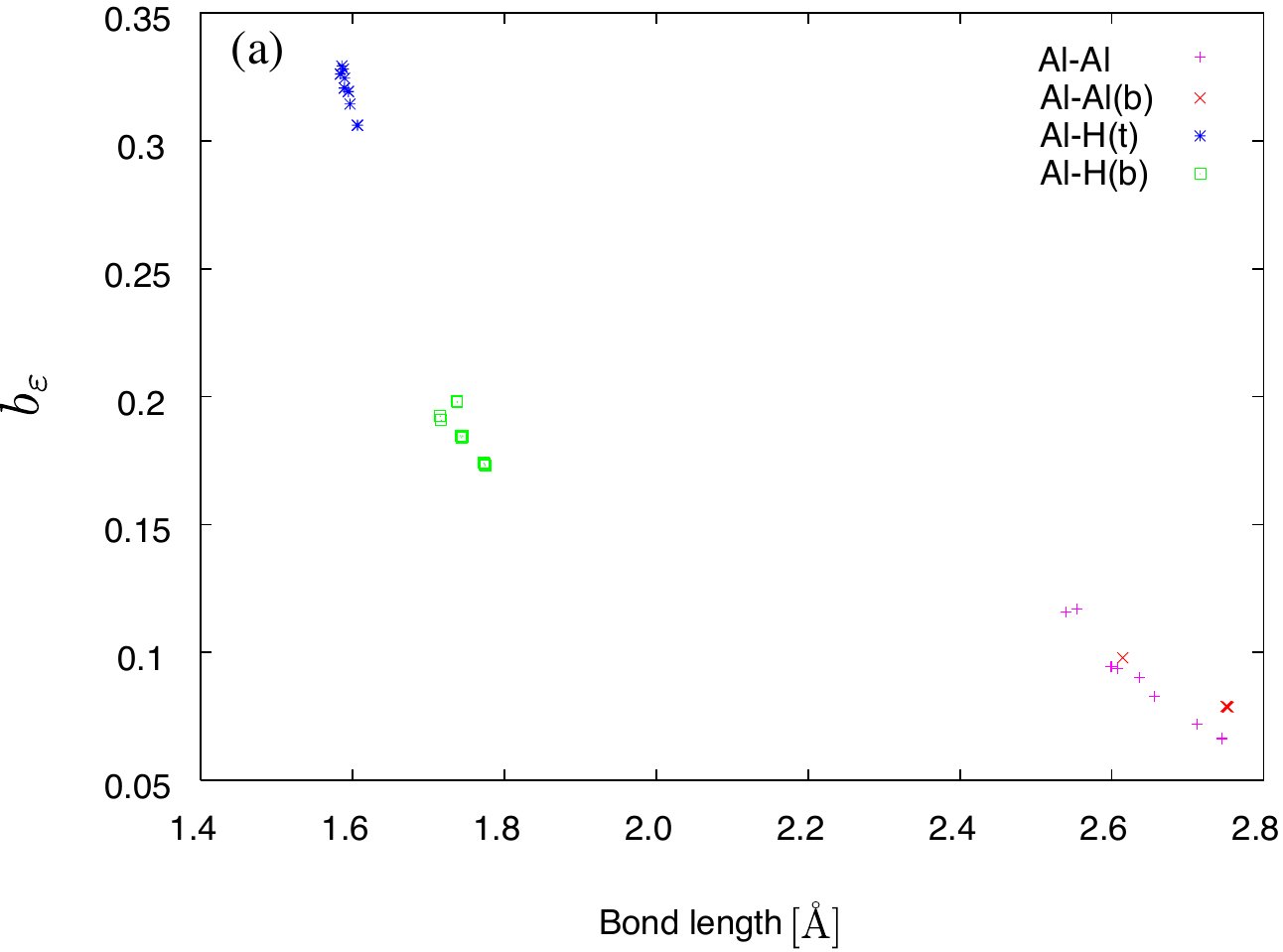}
\hfill
\includegraphics[scale=0.6]{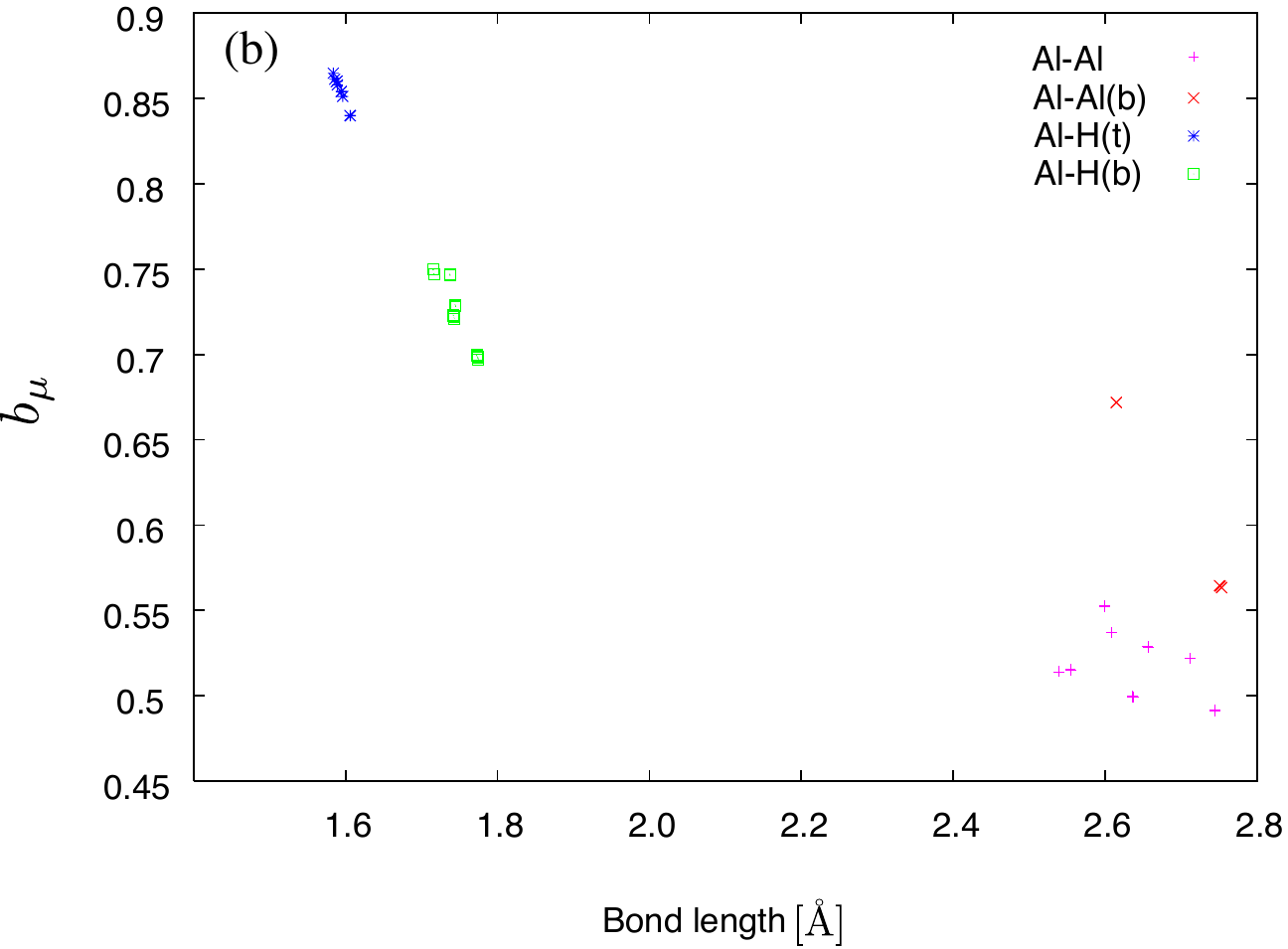}
\caption{Energy density bond order $b_\varepsilon$ (panel (a)) and chemical potential bond order $b_\mu$ (panel (b)) as functions of bond length. Data are taken from the structures of Al$_4$H$_n$ ($n=0, 2, 4, 6, 8, 10$ and 12) as shown in Fig.~\ref{fig:bo}. Al--Al bonds with bridging hydrogen are plotted with red crosses and those without are plotted with magenta plus marks. Al--H bonds at terminal sites are plotted with blue asterisks and those at bridge sites are plotted with green squares.}
\label{fig:dist_bo}
\end{center}
\end{figure}

Actually, we have already used the RDFT concept to draw Fig.~\ref{fig:bo}. There, we draw bond lines when the Lagrange point is found between two atoms.  The Lagrange point is the point at which tension density vanishes (Sec.~\ref{sec:RDFT}) and considered to be suitable to define chemical bond \cite{Szarek2007,Szarek2008,Szarek2009a}. 
Two types of bond orders, $b_\varepsilon$ (Eq.~\eqref{eq:be}) and $b_\mu$ (Eq.~\eqref{eq:bmu}), are computed and summarized as functions of bond distance in Fig.~\ref{fig:dist_bo}. The $b_\varepsilon$ describes bond strength in relation to bond in H$_2$ molecule and how particular bond contributes to lowering of total energy of a structure, whereas $b_\mu$ shows bond electrophilicity in relation to H$_2$ molecule as reference bond. 

In the figure, we distinguish different types of bonds. Al--Al bonds with and without bridging hydrogen are denoted by ``Al--Al(b)" and ``Al--Al". Al--H bonds at terminal sites are denoted as ``Al--H(t)" and those at bridge sites are denoted as ``Al--H(b)". We find good correlation between bond distance and our bond orders as has been found in other molecules \cite{Szarek2007,Szarek2008,Szarek2009a}. $b_\varepsilon$ and $b_\mu$ basically exhibit similar features regarding the correlation between bond length in the sense that the slope for Al--Al bonds are larger than Al-H bonds. There is a subtle difference in Al--H bonds between $b_\varepsilon$ and $b_\mu$. For $b_\varepsilon$, Al-H(t) and Al-H(b) together make a single slope but there seems to appear three families of slopes for $b_\mu$. Since $b_\mu$ concerns the regional chemical potential, the difference may reflect the chemical reactivity of these bonds.   

\begin{figure}
\begin{center}
\includegraphics[scale=1.2]{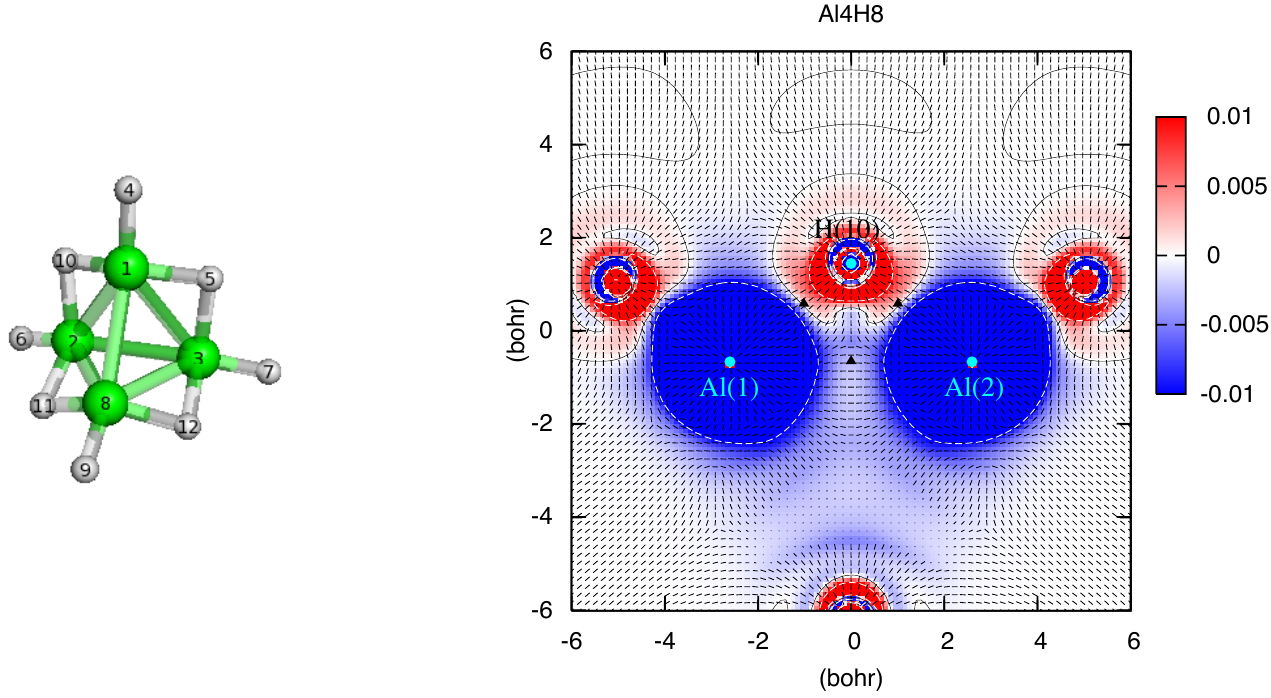}
\caption{The largest eigenvalue of the stress tensor and corresponding eigenvector of Al$_4$H$_8$ on the right panel. They are shown on a plane which includes three labeled atoms. As for the eigenvectors, the projection on this plane is plotted. The positions of these atoms are shown by the circle dots. Parenthesized numbers in the labels correspond to the numbers on atoms on the left panel. As for the eigenvalue, we only show for the range $[-0.01, 0.01]$ with color scale shown on the right and the contours for 0.01 and $-0.01$ are shown by white dashed lines. The triangle dots shows the locations of the Lagrange points. }
\label{fig:stress_Al4H8}
\end{center}
\end{figure}

\begin{figure}
\begin{center}
\includegraphics[scale=1.2]{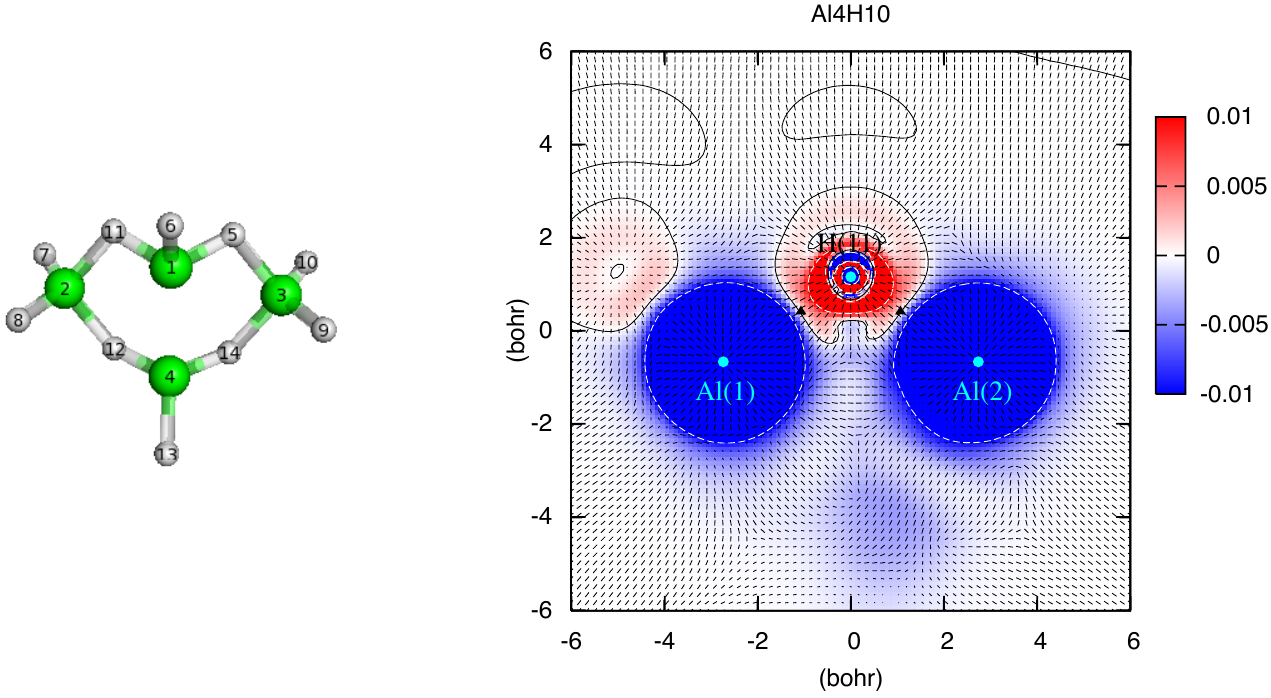}
\caption{The largest eigenvalue of the stress tensor and corresponding eigenvector of Al$_4$H$_{10}$, plotted in the same manner as Fig.~\ref{fig:stress_Al4H8}. }
\label{fig:stress_Al4H10}
\end{center}
\end{figure}

\begin{figure}
\begin{center}
\includegraphics[scale=1.2]{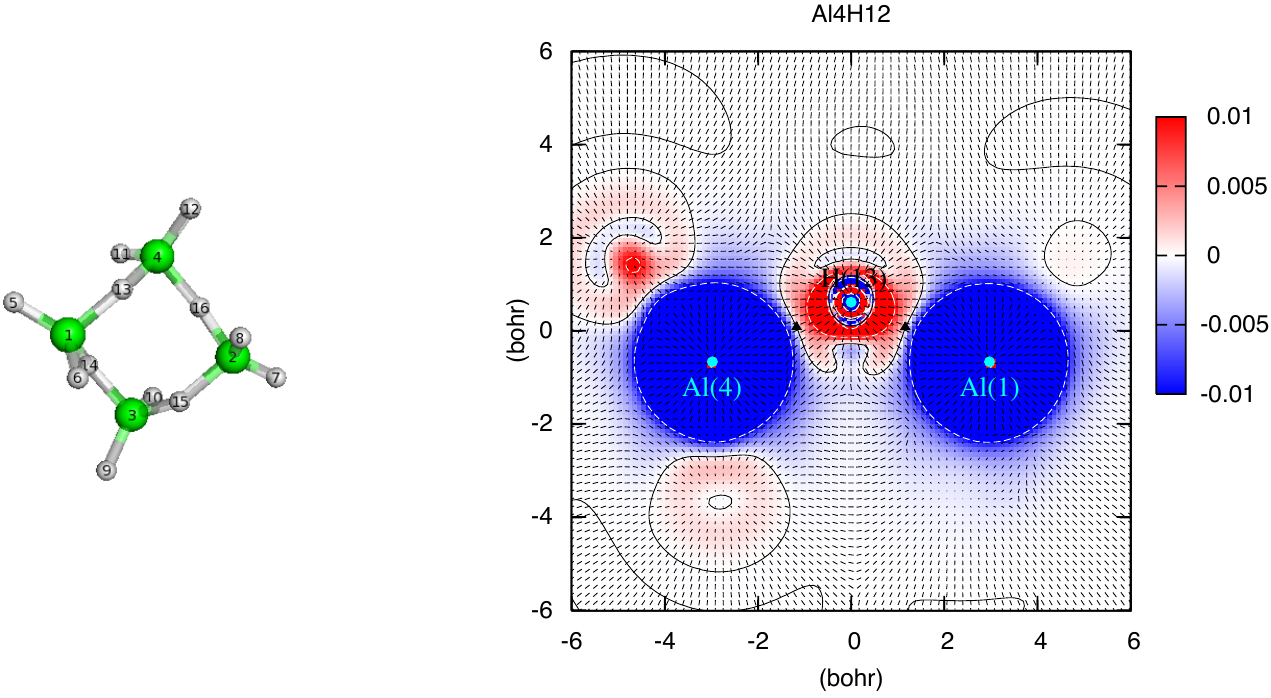}
\caption{The largest eigenvalue of the stress tensor and corresponding eigenvector of Al$_4$H$_{12}$, plotted in the same manner as Fig.~\ref{fig:stress_Al4H8}. }
\label{fig:stress_Al4H12}
\end{center}
\end{figure}

We can further examine features in chemical bonds by analyzing the electronic stress tensor (Eq.~\eqref{eq:stress}). For example, hydrogen bridged Al--Al bonds are investigated for Al$_4$H$_8$, Al$_4$H$_{10}$ and Al$_4$H$_{12}$. Figs.~\ref{fig:stress_Al4H8}, \ref{fig:stress_Al4H10} and \ref{fig:stress_Al4H12} plot the largest eigenvalue of the stress tensor and corresponding eigenvector on a plane which includes three atoms which participate in the bridging bonds, respectively for Al$_4$H$_8$, Al$_4$H$_{10}$ and Al$_4$H$_{12}$. 
Since the eigenvectors have three spatial components, namely they are 3D objects, we express them by projecting on the plane.
In all these clusters, there are Lagrange points between Al and H. There is a region with positive eigenvalue of the stress tensor (tensile stress) between them, which is typical for covalent bond involving H atom \cite{Tachibana2004,Ichikawa2009b}. Also, there is a flow of corresponding eigenvectors connecting Al and H which indicates a formation of strong bonding. 

For Al$_4$H$_8$, which has a Lagrange point between Al atoms, also has a flow of eigenvectors between them. In contrast to the case of Al-H bond, the eigenvalue of the stress tensor turns out to be negative (compressive stress) indicating a different nature of bonding. In the case of Al$_4$H$_{10}$, we did not find a Lagrange point between Al atoms. In Fig.~\ref{fig:stress_Al4H10}, there is a similar flow of eigenvectors as in Fig.~\ref{fig:stress_Al4H8} between Al atoms, but this structure is shifted off the region between Al atoms likely to be because the H atom came close to Al atoms. Therefore we may conclude there is not so much direct interaction between Al atoms as to call bonding. We can reach similar conclusion for the case of Al$_4$H$_{12}$. Fig.~\ref{fig:stress_Al4H12} shows similar structure but it is further away from the region between Al atoms and in fact (when we see the flow in 3D) the flow is connected to H(11) atom which locates at slightly off this plane. This indicates that Al--Al bonding is completely disrupted by the existence of H atoms. 

\begin{figure}
\begin{center}
\includegraphics[scale=1.2]{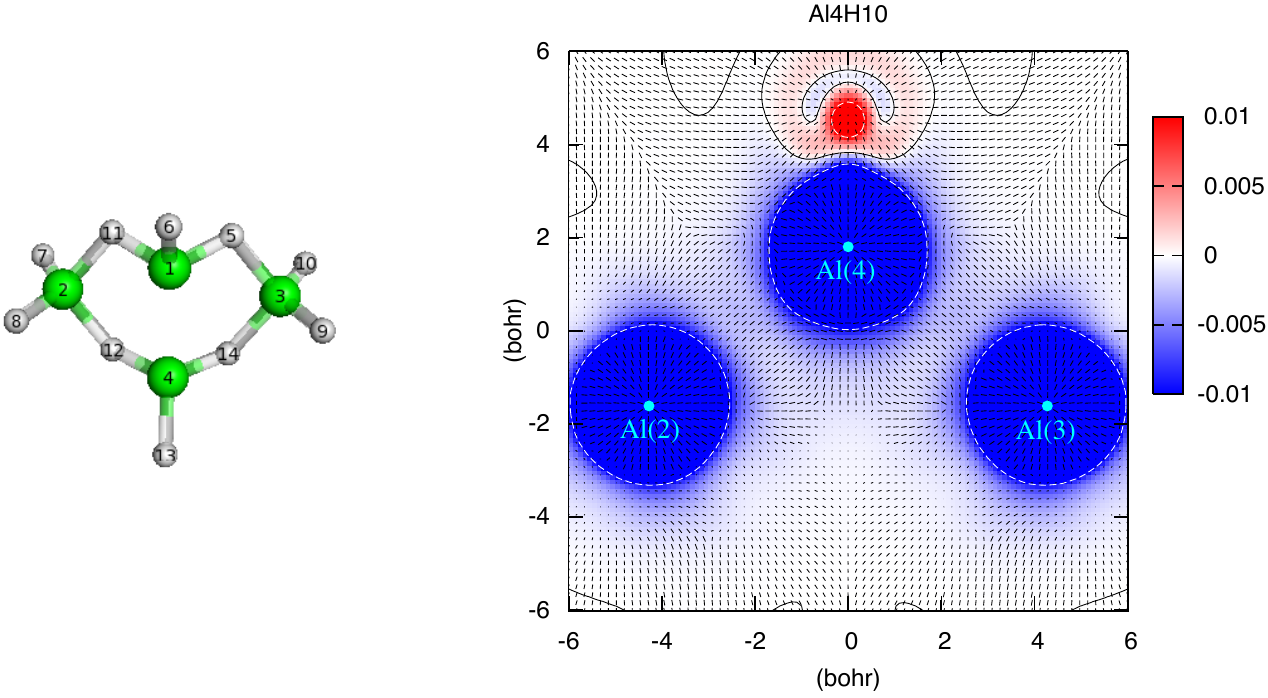}
\caption{The largest eigenvalue of the stress tensor and corresponding eigenvector of Al$_4$H$_{10}$ (but on the different plane from Fig.~\ref{fig:stress_Al4H10}), plotted in the same manner as Fig.~\ref{fig:stress_Al4H8}. }
\label{fig:stress_Al4H10_2}
\end{center}
\end{figure}

Another example of the electronic stress tensor analysis is provided for Al$_4$H$_{10}$ in Fig.~\ref{fig:stress_Al4H10_2}. This concerns the somewhat radical structural change between $n=0, 2, 4, 6, 8$ and $n=10, 12$ of Al$_4$H$_n$. As shown in Fig.~\ref{fig:bo}, in terms of Lagrange points, four Al atoms form a tetrahedral cage for $n \le 8$ and but the cage seems to be broken for $n \ge 10$. This is also seen in a jump in the average Al-Al distance (Table \ref{tab:Al4Hn}). We can confirm this by analyzing the stress tensor density. As is shown in Fig.~\ref{fig:stress_Al4H10_2}, there is a region between Al(2) and Al(3) in which eigenvectors go perpendicular to the plane (so that expressed by dots), showing total disconnection of these atoms.

\section{Summary} \label{sec:summary}
In this paper, we investigated the structures of aluminum hydrides derived from a tetrahedral aluminum (Al$_4$) cluster using {\it ab initio} quantum chemical calculation. 
We reported stable structures of Al$_4$H$_n$ ($n=0, 2, 4, 6, 8, 10$ and 12), which include structures already found in the literature who had investigated the hydrogenated aluminum clusters from other aspects.
We calculated binding energies of the aluminum hydrides and found interesting properties as hydrogen storage material: 
stability of the clusters increases as more hydrogen atoms are adsorbed, while stability of Al--H bonds decreases.

We also analyzed and discussed the chemical bonds of those clusters by using the electronic stress tensor. 
The bond orders defined from energy density and regional chemical potential (which are in turn calculated from the stress tensor) are shown to have a good correlation with respect to the bond distance and
to be able to distinguish types of bonding to some extent. As far as metallic elements are concerned, our bond order analysis had been only applied to Pt clusters \cite{Szarek2009a} so the present analysis can be useful basis for further research using our stress tensor based analysis. This is also true for the eigenvalue and eigenvector analysis of the stress tensor. We have found that Al--H bonds have a positive eigenvalue (tensile stress) at the region between the atoms where as Al--Al bonds have a negative value. This indicates that the stress tensor can be a powerful tool to classify chemical bonding and may provide a deeper insight into the nature of chemical bonds.


\newpage

\begin{table*}
\caption{Total B.E. $\Delta E_{\rm total}$, average B.E. of H atoms $\Delta E_{\rm H}$ and mean nearest-neighbor bond lengths $d_{\textrm{x-x}}$ of Al$_4$H$_n$ ($n=2, 4, 6, 8, 10$ and 12). H(t) and H(b) denote a hydrogen at the terminal site and the bridge site respectively.}
\begin{center}
\begin{tabular}{l ccccc}
\hline
\hline
& $\Delta E_{\rm total}$ (eV) & $\Delta E_{\rm H}$ (eV) & $d_{\textrm{Al-Al}}$ (\AA) & $d_{\textrm{Al-H(t)}}$ (\AA) & $d_{\textrm{Al-H(b)}}$ (\AA)  \\
\hline
 Al$_4$ & $-$5.59 & --- &2.74 & ---  & --- \\ 
 Al$_4$H$_2$ & $-$11.5 & $-$2.94 &2.66 & 1.61 & --- \\ 
 Al$_4$H$_4$ & $-$17.3 & $-$2.94 & 2.60 & 1.61 & --- \\ 
 Al$_4$H$_6$ & $-$22.7 & $-$2.84 & 2.63 & 1.60 & 1.74  \\
 Al$_4$H$_8$ & $-$27.6 & $-$2.75 & 2.69 & 1.59 & 1.77  \\
 Al$_4$H$_{10}$ & $-$32.2 & $-$2.66 & 3.11 & 1.59 & 1.74 \\
 Al$_4$H$_{12}$ & $-$36.6 & $-$2.59 & 3.47 & 1.59 & 1.72 \\
\hline
\end{tabular}
\end{center}
\label{tab:Al4Hn}
\end{table*}%

\newpage

\end{document}